\documentclass[doublecol]{epl2}

\title{Dynamics of particles and cages in an experimental 2D glass former}

\author{S. Mazoyer\inst{1} \and F. Ebert\inst{1} \and G. Maret\inst{1} \and P. Keim\inst{1}}
\shortauthor{S. Mazoyer \etal}

\institute{
  \inst{1} Fachbereich Physik, Universit\"at Konstanz, Universit\"atsstrasse
10, 78457 Konstanz, Germany.
}

\abstract{
We investigate the dynamics of a glass forming 2D colloidal mixture
and show the existence of collective motions of the particles. We introduce a mean square displacement $MSD$ with respect to the nearest neighbors which shows remarkable deviations from the usual $MSD$ quantifying the individual motion of our particles. Combined with the analysis of the self part of the Van Hove function this indicates a coupled motion of  particles with their cage as well as intra cage hopping processes.
}

\begin{document}

\maketitle

\section{Introduction}

Supercooled fluids near the glass transition exhibit a range of interesting dynamical properties such as
non-exponential relaxation functions, two time relaxation of the system or a dramatic increase of the time scale for molecular motion close to the glass transition ~\cite{Ediger1996}.
Most of these features have been attributed to spatially heterogeneous relaxation ~\cite{Berthier2005, Kob1997, Garrahan2005, Konig2005} and cooperative motion of particles ~\cite{Cui2001,Donati1998,Russell2000, Appignanesi2006}.
For instance confocal microscopy in a colloidal supercooled fluid have evidenced the existence of
populations of fast and slow particles, forming clusters of a few tens of fast colloids ~\cite{weeks2000}. Most of these clusters are visible only on the time scale of the order of the $\alpha$-relaxation.

Further common features of the dynamics of supercooled fluids are the behavior of
the self part of the Van Hove function and the non gaussian parameter ~\cite{Zahn2000, Odagaki1991,Donati1998, Szamel2005}.
Both of them reflect the non brownian character of particle motion and it is commonly accepted that the
maximum of the non gaussian parameter corresponds to a maximum in the heterogeneity of the dynamics.
More interesting maybe is the shape of the self part of the Van Hove function
~\cite{VanHove1954} from which more quantitative information can be extracted about the nature
of the motion.
The existence of two dynamical populations was confirmed from this quantity in numerous systems like granular media, colloidal gels, and Lennard Jones mixtures ~\cite{Dauchot2005,Chaudhuri2007, Appignanesi2006} and this behavior is also presented as a possible universal feature of glass forming systems ~\cite{Chaudhuri2007}.
Intensive studies of dynamical heterogeneities have been performed by simulations, while experimental
works in direct space remain quite rare ~\cite{Cui2001,weeks2000}. In this paper we present
results from a study of a 2D experimental colloidal system which consists of a binary mixture of
superparamagnetic particles interacting via a dipole-dipole interaction. This system allows to study the very nature
of the glass transition in 2D. In addition some of the local features like geometrical frustration are easier to detect compared to 3D ~\cite{Ebert2008,Ebert2009b,Tanaka2007,Harrowell2006}.

The average glassy dynamics of this system has been studied in ~\cite{Konig2005,Bayer2007}. Here we focus on the
microscopic local features of planar dynamics.

The organization of the paper is as follows. First we will briefly describe the experimental system. Second we will present experimental results and analysis of the dynamics for both the fluid phase and the supercooled phase.

\section{Materials and Methods}
\label{MaterialsandMethods}

The experimental setup is well established and has been described elsewhere
~\cite{Konig2005,Ebert2009}.

\begin{figure}
\includegraphics [width=0.48\textwidth]{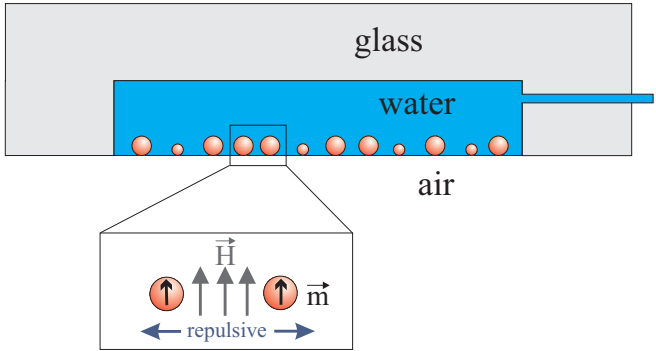}
\caption{\em Super-paramagnetic colloidal particles confined at a water-air
  interface due to gravity. The curvature of the interface is
  actively controlled to be completely flat, and the system
  is considered to be ideally two dimensional. A magnetic field $\textbf{H}$ perpendicular to the interface
  induces a magnetic moment $\textbf{M}$ in each bead leading to a repulsive dipolar
  pair interaction.}
\label{fig:ExpSetup}
\end{figure}

The system consists of a suspension of two kinds of spherical super-paramagnetic colloidal particles
A and B with different diameters ($d_A=4.5\,\mu m$, $d_B=2.8\,\mu m$) and magnetic susceptibilities per particle
($\chi_A\approx 10\cdot \chi_B$). Due to their high mass density of $\rho_m \approx 1.5\,g/cm^3$, particles are
confined by gravity to a water-air interface formed by a pending water drop suspended by surface tension
in a top sealed cylindrical hole ($6\:mm$ diameter, $1\:mm$ depth) in a glass plate. This basic setup is
sketched in \mbox{Figure ~\ref{fig:ExpSetup}}. A magnetic field $\textbf{H}$ is applied perpendicularly to the
water-air interface inducing a magnetic moment $\textbf{M}= \chi \textbf{H}$ in each particle leading to
a repulsive dipole-dipole pair interaction.\\
The parameter $\Gamma$ quantifies the strength of the interaction and is defined by the ratio between average
magnetic interaction energy and thermal energy.
$$\Gamma=\frac{\mu_0}{4 \pi}\frac{H^2.(\pi \rho)^{(3/2)}}{k_B T}(\xi . \chi_B + (1-\xi). \chi_A)^2$$
where $\xi$ denotes the relative number of small particles and $\rho$ is the 2D density.
$$\xi=\frac{N_B}{N_A+N_B}$$
The set of particles is visualized by video microscopy from
below the sample and is recorded by an 8-bit CCD camera. The gray scale image of the particles is then
analyzed \textit{in situ} with a computer. The field of view has a size of $\approx1\,mm^2$ containing
typically $3\times10^{3}$ particles, whereas the whole sample contains about up to $10^{5}$ particles.
Standard image processing is performed to
get size, number, and positions of the colloids. A computer controlled syringe driven by a micro stage
controls the volume of the droplet to reach a completely flat surface. To achieve a horizontal interface,
the inclination of the whole experimental setup has to be aligned. This inclination is controlled actively
 by micro-stages with a resolution of $\Delta\alpha \approx 1\:\mu$rad. After typically several weeks of
adjustment and equilibration best equilibrium conditions for long-time stability are achieved. During
data acquisition the images are analyzed with a frame rate down to $10\,Hz$. Trajectories of all particles
in the field of view can be recorded over several days providing the whole phase space information. The
thermal activated 'out of plane' motion of the particles is expected to be in the range of a few tens of
nanometer. Thus, the ensemble is considered as ideally two dimensional.\\ Information on all relevant time
 and length scales is available, an advantage compared to many other experimental systems. Furthermore, the
 pair interaction is not only known but can also be directly controlled over a wide range.

\section{Experimental observations}
\label{Experimentalobservations}

The study of the mean square displacement for this system has been presented earlier ~\cite{Bayer2007,Konig2005}. Here we only recall the main findings. At $\Gamma=25$ the system
is in a fluid state and the mean square displacement is diffusive at all time for both small and big particles.  For $\Gamma=110$ the system is in an intermediate phase, where the mean square displacement
exhibits an inflexion point around $t=1000$ $s$. For higher $\Gamma$, \textit{e.g} $\Gamma=338$ and $\Gamma=390$ the system is in a glass forming phase and the mean square displacement has 3 clearly distinct regimes. At early times, during the commonly called $\beta$-relaxation, it is diffusive. Then follows a plateau regime where the mean square displacement is almost constant. And finally one observes again an increase of the $MSD$, commonly called $\alpha$-relaxation.

\begin{figure*}
  \begin{minipage}[t]{.48\linewidth}
   \centering
    \includegraphics [width=1\textwidth]{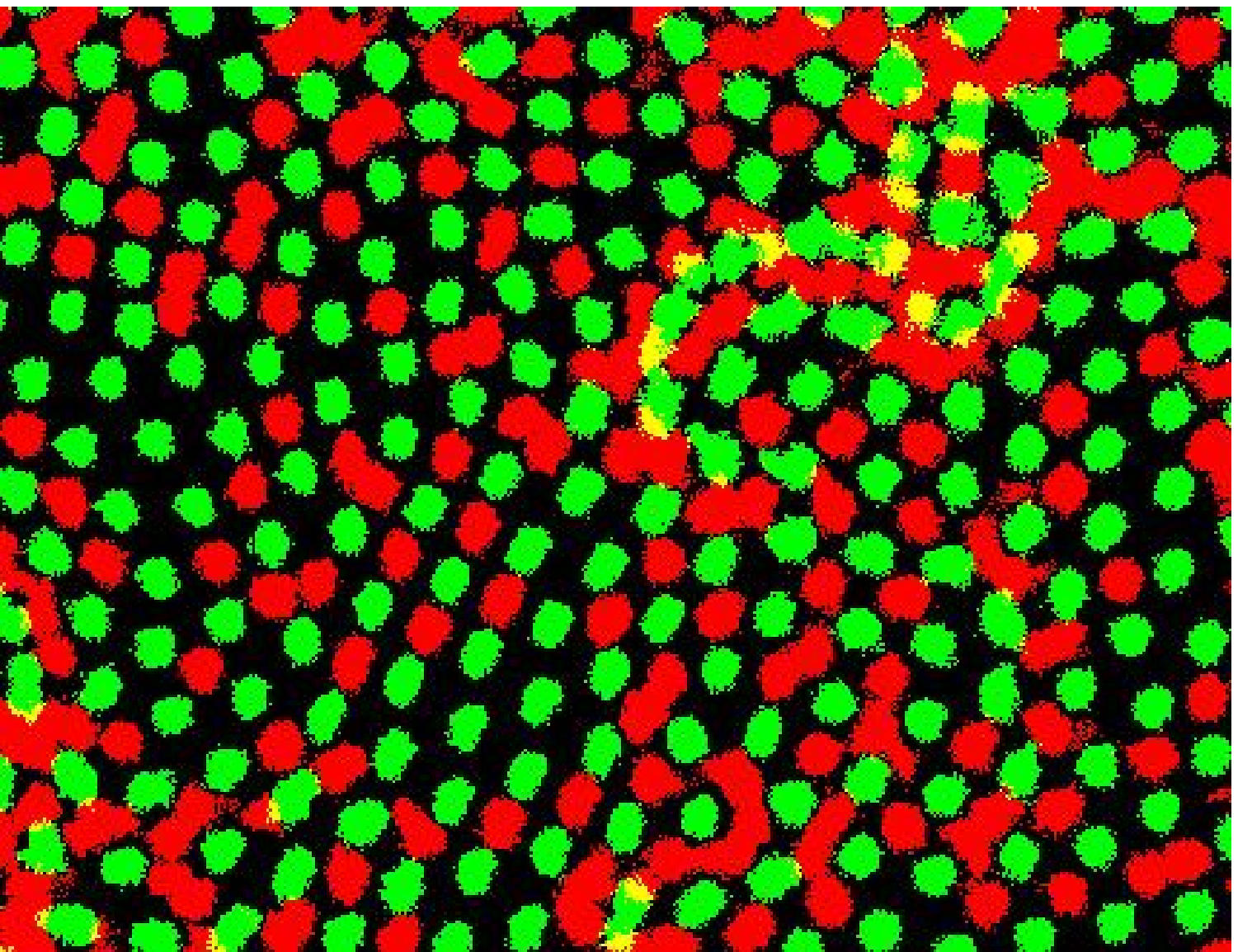}
  \end{minipage}
  \hfill
  \begin{minipage}[t]{.48\linewidth}
    \centering
    \includegraphics [width=1\textwidth]{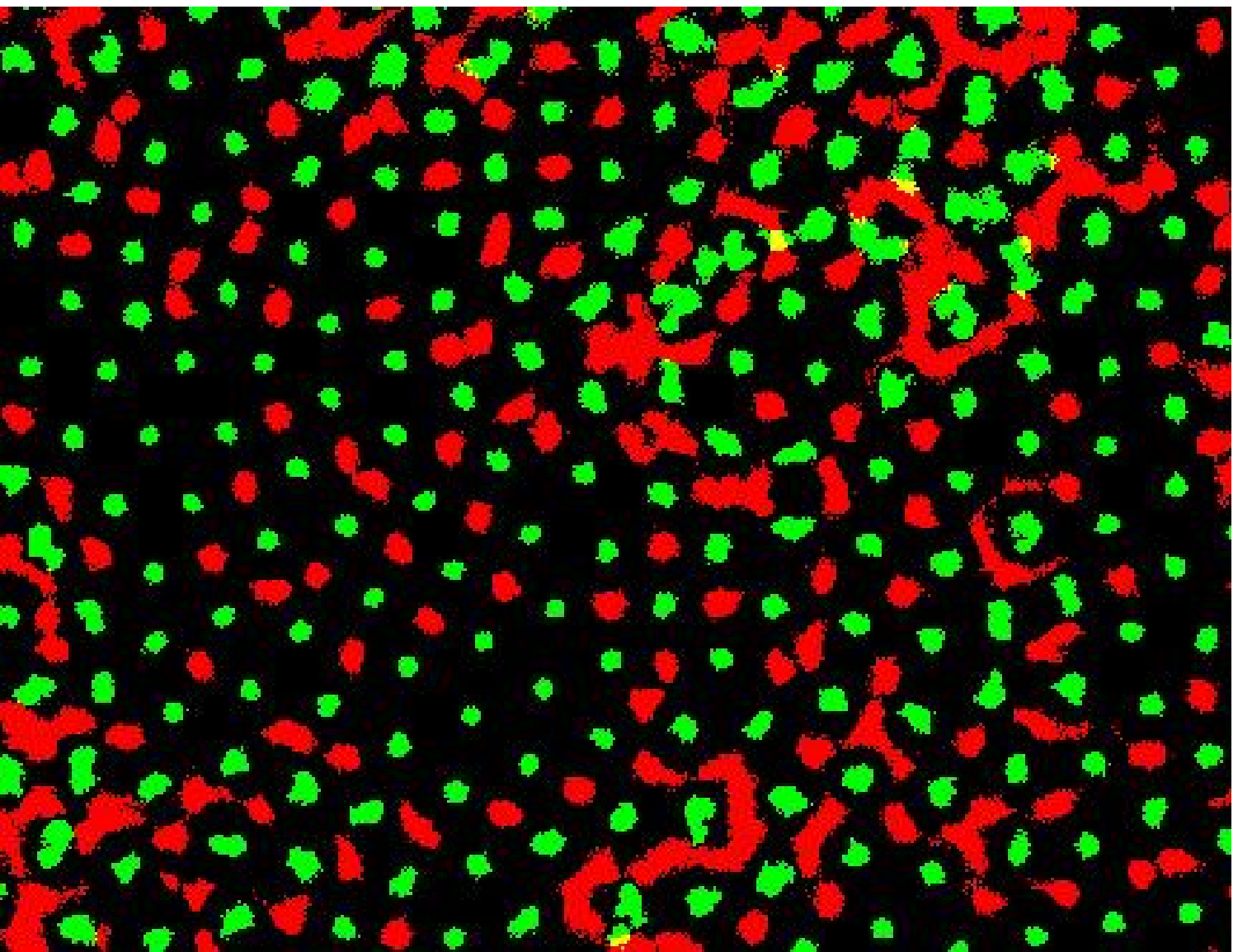}
  \end{minipage}
  \caption{a) Trajectories of big (in green) and small (in red) particles in the supercooled phase ($\Gamma=338$) within a box of $250 \times 180$ $\mu$m and over a duration of $t=80000$ $s$. b) Same as in a) but the trajectories of the particles are calculated relatively to the center of mass of their cage.}
  \label{fig:trajectories}
\end{figure*}

In order to get a better idea of microscopic dynamics we will address the question of caging of particles by their nearest neighbors. Particles escaping from their cage are often believed to be responsible for the $\alpha$-relaxation in the $MSD$. Therefore we investigate the displacement of a colloid with respect to the average displacement of its nearest neighbors as a function of time. We define the cage relative $MSD$ as follow :
\begin{equation}
\langle\Delta r^2_{CR}(t)\rangle =
\langle [(\vec{r}_i(t)-\vec{r}_i(0))-
(\vec{r}_i^{cage}(t)-\vec{r}_i^{cage}(0))]^2\rangle
\end{equation}
where $\langle \rangle$ is the ensemble average, $\vec{r}_i(t)$ is the position of the
particle $i$ at time $t$, and $\vec{r}_i^{cage}$ the position of the center of mass of the initially nearest
neighbors:
$\vec{r}_i^{cage}=\frac{1}{N_{nn}}\sum_{j=1}^{N_{nn}}(\vec{r}_j(t)-\vec{r}_j(0))$ where $j$ runs over the
indices of the nearest neighbors defined by Voronoi tessellation and $N_{nn}$ is the number of nearest neighbors.
The cage relative $MSD$ was successfully used in simulations ~\cite{Earnshaw1998} and experiments ~\cite{Zahn2001} of crystallizing systems to determine the melting temperature in 2D. For a fluid system the cage relative $MSD$ diverges as function of time whereas it saturates in the crystalline, arrested state.
In Fig.~\ref{fig:trajectories} we have plotted trajectories of particles for $\Gamma=338$ within a box of $250 \times 180$ $\mu$m and also their cage relative trajectories. The trajectories have been plotted for the entire duration of the experiment, i.e. 80 000 $s$, which is close to the $\alpha$-relaxation time for this value of $\Gamma$.\\

In Fig.~\ref{fig:trajectories} a) the dynamics appears strongly heterogeneous, with zones where trajectories are almost isotropic and some others where they are elongated, forming zones of fast moving particles. Both big and small particles are involved in such clusters. Dynamical heterogeneities are even more visible in the cage relative trajectories and cage relative trajectories are also more compact.
The existence of such compact clusters of fast moving particles is in agreement with what has been found in other systems like 3D colloidal glass ~\cite{weeks2000} or molecular glass formers ~\cite{Castillo2007, glotzer1999}. Presence of a few string-like motion has to be noticed, as in quasi 2D colloidal system ~\cite{Cui2001} or Lennard Jones mixtures ~\cite{Kob1997,Appignanesi2006} but they do not represent a significant part of the rearranging clusters.\\

\begin{figure}[h]
\includegraphics [width=0.45\textwidth]{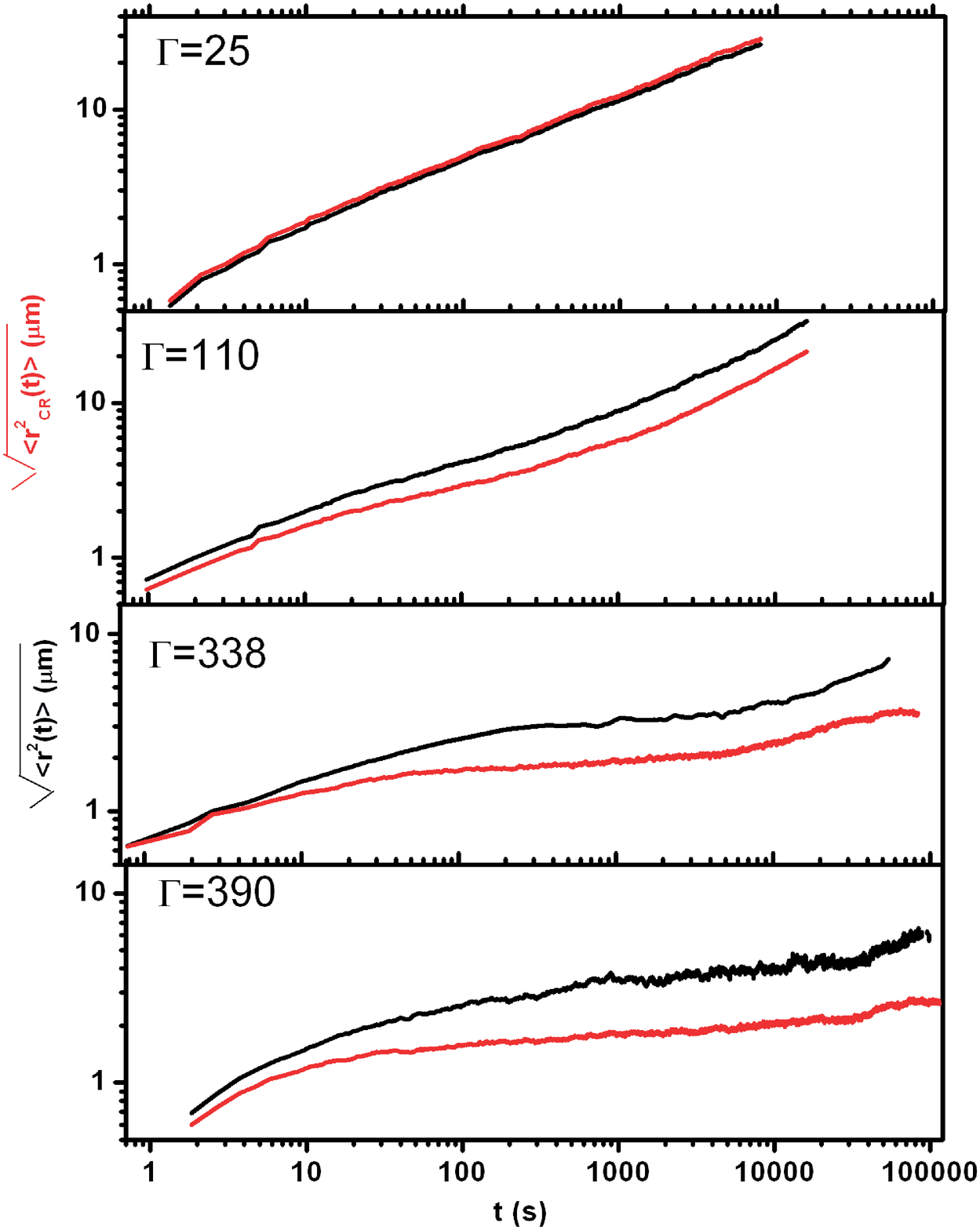}
\caption{Cage relative $MSD$ as a function of time, in red and usual $MSD$ in black,
both for  $\Gamma=25$, $\Gamma=110$, $\Gamma=338$ and $\Gamma=390$, with a Log-Log scale.}
\label{fig:lindemannparameter}
\end{figure}

In Fig.~\ref{fig:lindemannparameter} we have plotted both, the cage relative $MSD$ and $MSD$
for the fluid phase, $\Gamma=25$, the intermediate phase, $\Gamma=110$ and the supercooled case, $\Gamma=338$ and $\Gamma=390$.
Global drift of the system has been subtracted.
For the fluid phase at $\Gamma=25$, we see that $MSD$ and cage relative $MSD$ are very similar in shape
 and are very close to each other (with a ratio less than 1.1). Cage relative $MSD$ is slightly larger than $MSD$, which corresponds to the situation where the motion of neighboring particles is nearly uncorrelated (as expected in the fluid phase) and the heavy-center is pushed in the opposite direction if the center-particle leaves the cage (producing a kind of counterflow). In the intermediate phase, $\Gamma=110$, cage relative $MSD$ and classical $MSD$ are quite similar but cage relative $MSD$ is always smaller than $MSD$. The deviation becomes larger at the inflection point. Similarities in the curves of $MSD$ and cage relative
$MSD$ in the intermediate phase indicate that the motion of particles almost corresponds to an individual motion, uncorrelated with the motion of its neighbors, like in the fluid phase. The situation changes when we look at the curves for the
supercooled phase at $\Gamma=338$ and $\Gamma=390$. For early times curves are similar in shape and value, but
very quickly (after 10 s) the two curves start to significantly differ and cage relative $MSD$ remains significantly
smaller than $MSD$. For the longest times $MSD$ is even twice larger than the cage relative $MSD$.
This shows that in the supercooled phase the motion
of particle is of two types: firstly, an intra cage motion which is predominant over timescales of the
order of seconds and secondly, a motion of particles with the cage which starts to appear at the end of the $\beta$-relaxation of
the $MSD$ and is of the same order than the intra cage motion in the plateau. This significant collective motion of cages is a characteristic point of dynamical heterogeneities in our system for the supercooled phase. In order to analyze more deeply the nature of the motion of individual particles we investigate a quite usual
quantity which is the number of particles $N(r)$ which are at a distance $r$ from their original position
after a given time $t$. This quantity is expressed for various times as a function of $r$.
In 2 dimensions, $N(r)$ must be divided by the geometrical factor $2 \pi r$ to correspond to a probability and is called self part of the Van Hove function.
Brownian motion for instance, would give the characteristic gaussian shape for $\frac{N(r)}{2 \pi r}$.
Differences to gaussian behavior are usually associated to dynamical heterogeneities ~\cite{Appignanesi2006,Chaudhuri2007}. Here we normalize all curves to one at $r=0$ in order to facilitate comparison between the shapes of the curves.\\

\begin{figure}
\includegraphics [width=0.45\textwidth]{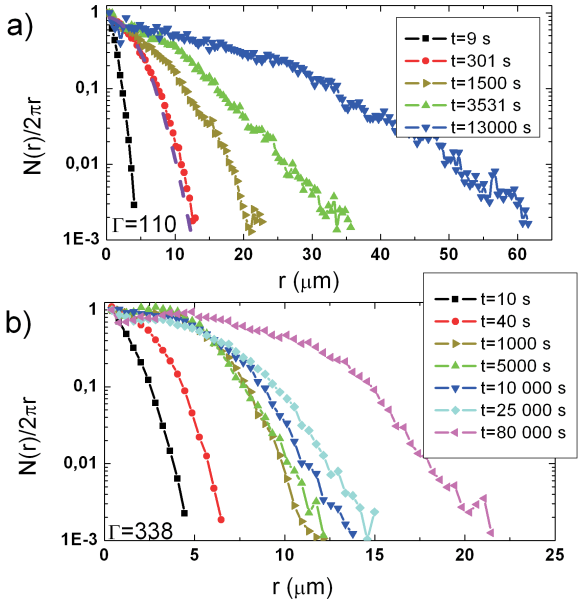}
\caption{a) Self part of the Van Hove function $\frac{N(r)}{2 \pi r}$ (after normalization) for $\Gamma=110$ for various times and with a Lin-Log scale. (Dashed line) gaussian fit of $\frac{N(r)}{2 \pi r}$ for $t=301$ $s$. b) Same quantity for $\Gamma=338$ and for various times.
Scale in x-coordinate is not the same for the two values of $\Gamma$}
\label{fig:VanHove}
\end{figure}

In Fig. ~\ref{fig:VanHove} we have plotted the quantity $\frac{N(r)}{2 \pi r}$ for the
 values of $\Gamma$ in the intermediate and supercooled phase respectively and for different times covering all
the regimes of the $MSD$. We have averaged over 10 successive times, starting at the time indicated as label of the curves.
The fluid case is trivial and the self part of the Van Hove function is a gaussian
(not shown here).
For $\Gamma=110$ and for early times the profile of the $\frac{N(r)}{2 \pi r}$ is gaussian as indicated by the good agreement with gaussian fit and does not differ very strongly from it at any time.
Distance performed by particles are much larger than inter particle distance ($L=23$ $\mu$m), so particles escape from the cage created by their nearest neighbors.

For $\Gamma=338$ the situation is different. Although the curve remains close to a gaussian at short times, deviations from gaussian start to appear in the plateau regime and later. Usually such deviations from gaussian behavior are associated with dynamical heterogeneities.\\ 
It seems quite natural to calculate what is the equivalent for the self part of the Van Hove function of the cage relative $MSD$.
This cage relative self part of the Van Hove function, $\frac{N_{CR}(R)}{2 \pi R}$, is defined by the number of particles at a relative distance $R$ from its origin in regard to the initial cage of the particle, expressed as a function of $R$ and normalized as previously. For a particle $i$ the relative distance $R$ to initial cage is defined as follow:
\begin{equation}
R_i=\mid(\vec{r}_i(t)-\vec{r}_i(0))-\Delta \vec{r}^{cage}_i\mid
\end{equation}
where $\Delta \vec{r}^{cage}_i$ is defined by $\frac{1}{N_{nn}}\sum_{j}^{N_{nn}} (\vec{r_j}(t)-\vec{r_j}(0))$. Here, $j$ runs over nearest neighbors, $N_{nn}$ the number of nearest neighbors, and $\vec{r}(t)$ the position of a particle
at time $t$. Like for the self part of the Van Hove function we have normalized all curves to have a maximum
equal to one.
\begin{figure}
\includegraphics [width=0.45\textwidth]{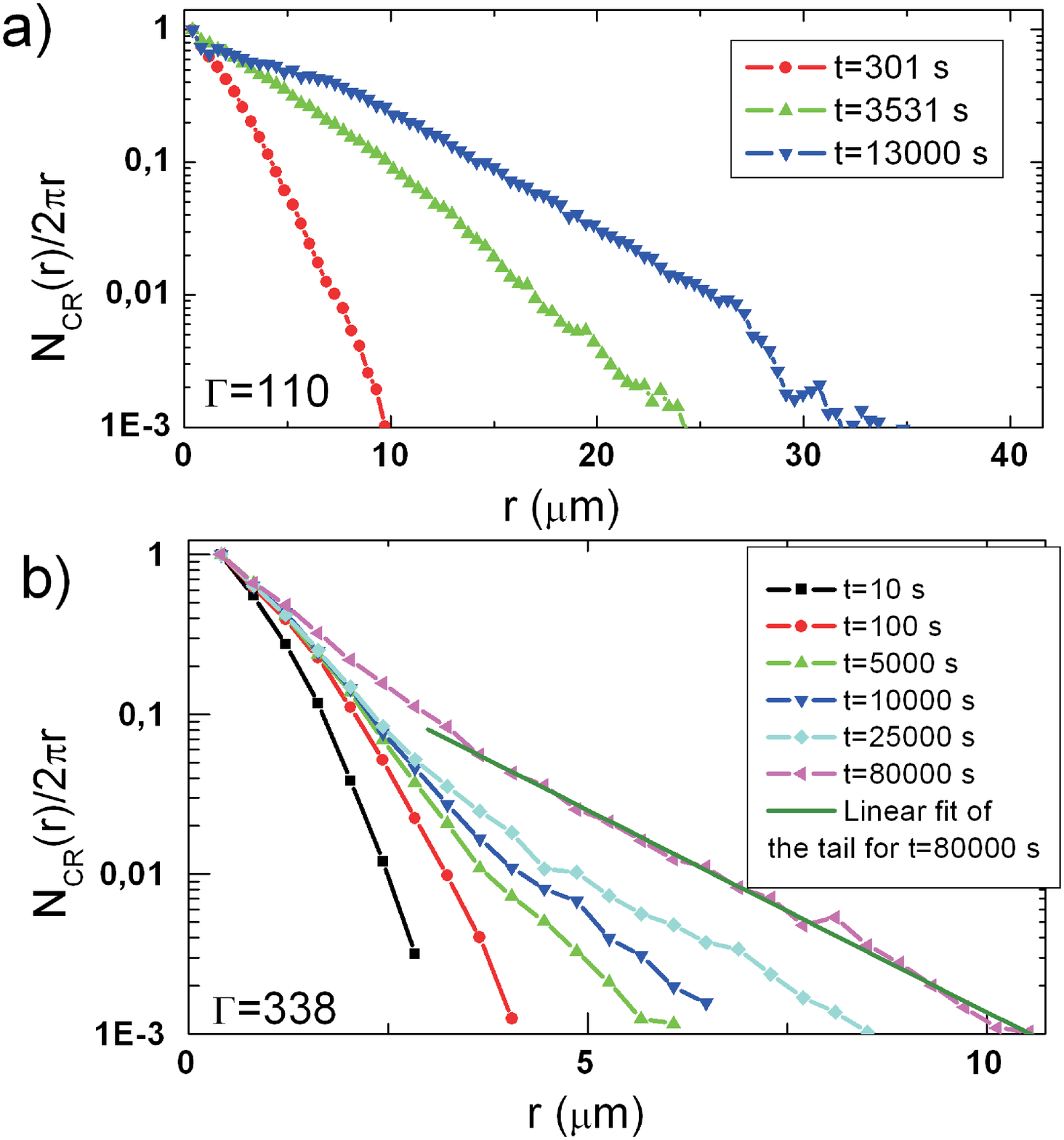}
\caption{a) Cage relative self part of the Van Hove function $\frac{N_{CR}(R)}{2 \pi R}$ (after normalization) for $\Gamma=110$ for various times and with a Lin-Log scale.
 b) Same quantities for $\Gamma=338$ and for different values of times.}
\label{fig:VHLindFig}
\end{figure}
In Fig. ~\ref{fig:VHLindFig} we have plotted $\frac{N_{CR}(R)}{2 \pi R}$ for the intermediate case ($\Gamma=110$) and the supercooled case ($\Gamma=338$) for big particles. All these quantities are averaged over 10 successive starting times.
For $\Gamma=110$, the curves are very similar in shape to those of $\frac{N(r)}{2 \pi r}$. Differences come only from the fact that the distribution is more narrow, which corresponds to the fluid like behavior of the cage relative $MSD$.
\\
For $\Gamma=338$, the situation is completely different. Before the plateau regime the curves look gaussian in shape, but their behavior changes at the beginning of the plateau ($t \approx 100$ $s$) when deviation from gaussian behavior occurs: the central part seems to remain mainly unchanged but a tail starts to appear. For large enough times the tail becomes exponential, as indicated by the linear fit in Lin-Log scale.

Compared with results from classical self part of the Van Hove function other systems, the behavior of its cage relative version now matches what is observed usually in 3D systems.
In ref. ~\cite{Chaudhuri2007}, Chaudhuri \textit{et al.} describe the exponential tail as a possible universal feature for jammed systems and supercooled fluids exhibiting dynamical heterogeneities. Common idea about this tail is that it corresponds to jumps of particles out of their cage. In our case, displacement of the particles in the tail remains largely lower than the average inter particle distance ($L=23$ $\mu$m), so jumping particles remain confined in their cage. This is also compatible with what has been previously seen in silica glass or Lennard Jones mixtures \cite{Chaudhuri2007}.

Another interesting fact is that in $\frac{N_{CR}(R)}{2 \pi R}$ for
$\Gamma=338$, the gaussian central parts seem to be very similar for
all times. We have checked (data not shown) that if we subtract
$\frac{N_{CR}(R)}{2 \pi R}$ taken at time $t=50$ $s$ from its
counterparts after $t=100$ $s$ we obtain in all cases curves
corresponding to a pure exponential decay (except at very low $r$ for
which uncertainty of the data is too high). For low temperatures the residual motion belonging to the gaussian part of the cage relative Van Hove function is that of a particle being trapped in a potential minimum created by its neighbors. The standard deviation of the gaussian probability distribution of positions of the particle increases soon (within the $\beta$-relaxation regime) towards an asymptotic value.  This nicely demonstrates the validity of the cage-picture of mode coupling theory. Any further increase of cage relative $MSD$ may correspond to some jumps (or hopping process) of the particles, indicated by the tail of the cage relative self part of the Van Hove function. This image is consistent with improvements of MCT theory developed to describe hopping processes (see \cite{Schweizer2003} for instance).\\
Differences with classical self part of the Van Hove function are
obvious: jumps of particles which may be hidden by collective motion are visible in the cage relative version.
This way, they are already visible in the plateau regime where collective unidirectional motion (included in the measure of the classical MSD) may hide them.\\

\begin{figure}
\includegraphics [width=0.48\textwidth]{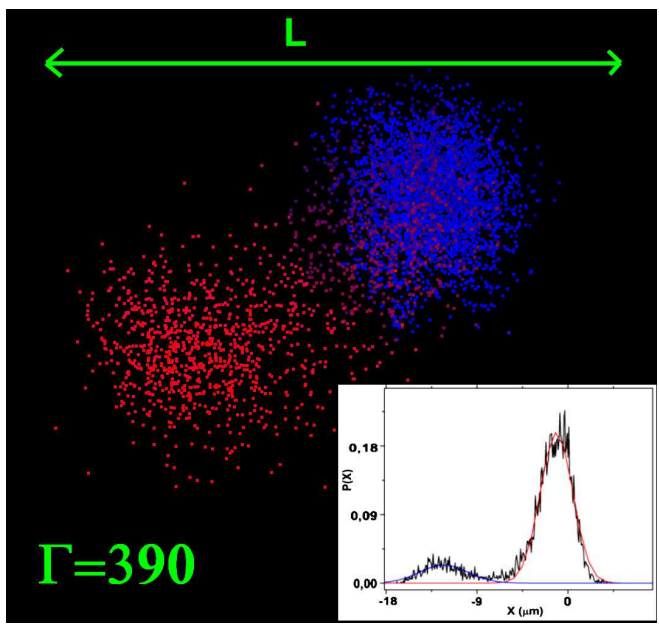}
\caption{Trajectory of a single big particle chosen amongst the 5\% fastest particles for a sample at
$\Gamma=390$. The color code ranges from pure blue for early times to pure red for the latest times. The inset shows the projection of the density of presence of the particle along the transversal axis of the trajectory. Red and blue curves are gaussian fits of the peaks. The line represents the average interparticle distance and corresponds to $L=23$ $\mu$m}
\label{fig:trajectory}
\end{figure}

In Fig.~\ref{fig:trajectory} a typical trajectory of a big particle for $\Gamma=390$ is plotted.
The particle belongs to the 5\% fastest particles and therefore to the tail of the cage relative self part of the Van
Hove function for $t=113000$ $s$.
It shows a clear intra cage hopping process. During early times
the particle remains confined around its initial position and explores the cage.
After this phase of exploration the particle performs a jump which takes about
$t_{jump} \approx 200$ $s$ and starts a new cage exploration around this position. The distance $d_{jump} \approx 10$ $\mu$m
performed during the jump is much smaller than the inter particle distance so the jump can not be
explained simply by a jump from one cage to another but must be more subtle
phenomenon.
This behavior was observed previously both in simulations ~\cite{Chaudhuri2007}, and experiments \cite{weeks2000,Marty2005}. As noticed in ref. ~\cite{Chaudhuri2007} and ~\cite{weeks2000}, the jump duration is very small compared to the time needed for a cage exploration. The nature of this jumping process is still under debate and many authors ~\cite{Chaudhuri2007,Appignanesi2006} invoke cooperative motion of the particles forming the cage to justify displacement smaller than the average inter particle distance.

\section{Conclusion}
\label{Conclusion}

In this work we have developed a new analysis tool to provide evidence of two different kinds of motion in an experimental 2D glass former. The use of the cage-relative mean square displacement (CR-MSD) allowed us to identify a typical cage dynamics of the particles and dynamical heterogeneities are much more pronounced.
In the short time limit, particles perform free diffusion until they start to feel the neighboring particles in the supercooled stage. In regard to the cage made by the nearest neighbors, particles behave like brownian particles in a potential minimum corresponding to the plateau in the $MSD$. Most particles remain blocked inside the cage, while a few of them start to make some hopping process already in the plateau regime. The tail of the cage relative self part of the Van Hove function, which corresponds to hopping processes becomes significant in the $\alpha$-relaxation regime.
But comparing the length scales of the inter-particle distance and the plateau hight of the $MSD$ one finds that most of the fast particles do not completely leave their neighborhood. This dynamical process, despite the fact that it does not present any large string motion as in Ref. ~\cite{Kob1997,Appignanesi2006}, may correspond to cooperative rearrangements which are seen in most 3D systems.
In addition to this, the difference between the classical $MSD$ and the cage relative $MSD$ has shown the presence of an important collective motion of particles with their cage which is especially large in the plateau and the $\alpha$-relaxation regime. This collective motion hides the contribution of the classical cage dynamics of the particles to the $MSD$ and, looking at trajectories of Fig.~\ref{fig:trajectories}, is expected to have characteristic length scale of several cage sizes.
\\
\acknowledgments
This work was supported by the DFG (Deutsche Forschungsgemeinschaft) in the frame of SFB TR6 project C2 and we thank Djamel EL Masri and Ludovic Berthier for fruitful discussion.

\end{document}